# Selective Near Perfect Light Absorbtion by Graphene Monolayer Using Aperiodic Multilayer Microstructures

Iman Zand[1,2,6], Hamed Dalir[3], Ray T. Chen[4], and Jonathan P. Dowling[1,2,5]

[1]*Center for Computation and Technology, Louisiana State University, Baton Rouge, Louisiana 70803, USA*

[2] *Hearne Institute of Theoretical Physics and Department of Physics and Astronomy, Louisiana State University, Baton Rouge, Louisiana 70803, USA*

[3]*Omega Optics, Inc., 8500 Shoal Creek Blvd., Bldg. 4, Suite 200, Austin, TX 78757, USA2*

[4]*Department of Electrical and Computer Engineering, The University of Texas at Austin, 10100 Burnet Rd., MER 160, Austin, TX 78758, USA*

[5]*New York University Shanghai, Shanghai Shi 200122, China.*

[6]*Hub Petroleum, 1100 Poydras St., Suite 2925, New Orleans, Louisiana 70163, USA*

*\*Corresponding authors: izand@hubpetroleum.com, hamed.dalir@omegaoptics.com*

We investigate 1D aperiodic multilayer microstructures in order to achieve near total absorption in preselected wavelengths in a graphene monolayer. Our structures are designed by a genetic optimization algorithm coupled to a transfer matrix code. Coupled mode theory (CMT) analysis, in accordance with transfer matrix method (TMM) results, indicates the existence of a critical coupling in a graphene monolayer for perfect absorptions. Our findings show that the near-total-absorption peaks are highly tunable and can be controlled simultaneously or independently in wide range of wavelengths in the near-infrared and visible. Our proposed approach is metal free and does not require surface texturing or patterning, and can be applied for other two dimensional (2D) materials.

Graphene as a 2D material (one-atom thick) has opened a new horizon for researchers due to its unique electronic and optical properties [1-5]. However, one of the critical issues for designing graphene-based optical devices is the enhancement of absorption of light in order to achieve near total absorption. In the mid- to far-infrared, various optical structures based on patterned graphene [6,7] and unpatterned graphene [8-10] have been introduced. These structures take advantage of the plasmonic response of graphene. On the other hand, in the visible and near-infrared regime, undoped, unpatterned graphene dose not exhibit a plasmonic response [11], and it only absorbs about 2.3% of the light at normal incidence [12]. In order to enhance light absorption in graphene monolayer plasmonic antennas [13], photonic crystal slabs [14-17], and Fabry-Perot microcavities [18-27] have been studied.

Among the proposed platforms, one dimensional (1D) dielectric multilayers are promising candidates because of their simplicity of design and low amount of optical loss compared to metallic structures. In previous works based on 1D multilayer structures, localization of light near a graphene monolayer has been achieved by placing graphene inside an asymmetric Fabry-Perot cavity, with a partially reflective front mirror and a perfect back mirror. Since in visible and near-infrared regimes undoped graphene does not have plasmonic response, coupling of light to graphene is governed by properties of resonance modes of 1D multilayer structures. Hence, spatial localization of longitudinal resonant modes of the multilayer structure directly influences the absorption spectrum of the graphene monolayer. One of the efficient approaches to control spatial localization of resonance modes in 1D structures is to use aperiodic multilayers supporting asymmetric/symmetric and extended/localized field profiles [28]. In fact, aperiodic structures, compared to asymmetric Fabry-Perot based designs, gives us more degrees of freedom and efficiency for the spatial selective localization of resonant modes. Although in asymmetric Fabry-Perots' front and back mirrors are composed of periodic layers, in aperiodic multilayers both sides of the graphene monolayer are non-periodic multilayers with arbitrary layers' thicknesses.

Hence, due to importance of applications requiring multispectral light detection, in this contribution, aperiodic multilayers are proposed as efficient platforms to enhance (~100% absorption) and tune light absorption in a graphene monolayer at multiple preselected wavelengths. Here, we optimize thicknesses using a genetic optimization algorithm [29-31], coupled to a transfer matrix code, in order to maximize absorption of the graphene monolayer at desired preselected wavelengths. In addition, CMT is used to explain physics behind the resonating structures. Our simulation results show that proposed approach not only provides near total light absorption at preselected multiple wavelengths but also help us to efficiently tune these wavelengths either simultaneously or independently.

A schematic of the aperiodic design is shown in Fig. 1. As seen, a graphene monolayer is placed inside of 1D aperiodic multilayer composed of alternating layers of silicon and silica deposited on a silica substrate. The two adjacent layers of the graphene monolayer are set to be $SiO_2$ in order to have homogeneous environment surrounding the graphene monolayer. We model optical properties of silicon and silica using modeled by the experimental data of Ref. [32], and the graphene monolayer is characterized as in Ref. [14, 15]. The first layer of the top (bottom) multilayer is silicon (silica) and the last layer is silica (silicon). For simplicity, we refer to structures as $M_iN_j$, where M and N stand for top and bottom multilayers, and $i$ and $j$ are number of layers. Top multilayer consists of 6 layers and the bottom multilayer consists of 16 or 18 layers. All the layers are allowed to vary between 40 nm and 600 nm. Our simulation results are based on the light incident from air into the structures.

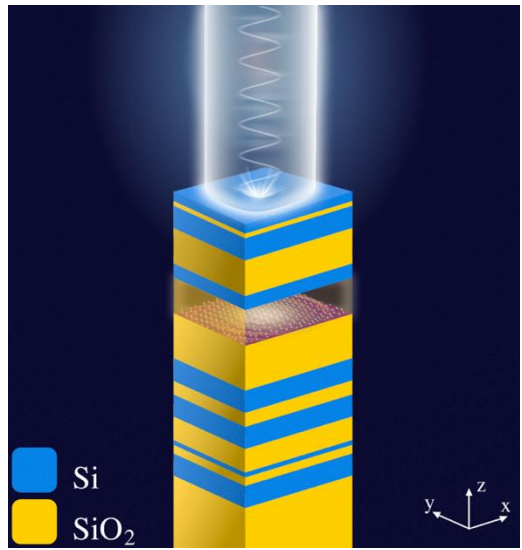

Fig. 1. Schematic of a one-dimensional aperiodic multilayer composed of alternating layers of silicon (blue) and silica (yellow) on top of a semi-infinite silicon substrate. The graphene monolayer (red) is placed inside of the structure.

Our main goal is to design 1D aperiodic multilayers to achieve highly tunable near total absorption by a graphene monolayer at preselected wavelengths. We take advantage of hybrid optimization algorithm in combination with transfer matrix method to maximize the absorption in the graphene monolayer at multiple preselected wavelengths. The optimization algorithm consists of micro-genetic global optimization algorithm coupled to an optimization suite packaged by Massachusetts Institute of Technology called NLOPT [29-31]. This algorithm enables us to determine the best structures' dimensions for varying number of layers at a given wavelength, $\lambda_i$. In order to gain maximum absorption at several wavelengths, we maximize the fitness function, $F = \sum_{i=1}^{n} A(\lambda_i)$, where $A(\lambda_i)$ is the absorption in graphene monolayer at wavelength $\lambda_i$.

To calculate optical power absorbed in a graphene monolayer, we first calculate the total reflection (R) of the aperiodic structure to obtain the field amplitude of the reflected wave. Then field amplitudes in adjacent layers of graphene can be determined by matrix multiplication of the individual components of the layers, starting from the first region (which is air) and proceeding to the adjacent layers. Once the forward and backward fields in vicinity of graphene are known, their corresponding Poynting vectors lead us to compute optical power absorbed by the graphene monolayer [19]:

$$A(\lambda, \theta) = S_{\text{in}}(\lambda, \theta) - S_{\text{out}}(\lambda, \theta) \qquad (1)$$

where $A$ is absorbance, $S_{\text{in}}$ ($S_{\text{out}}$) is the optical power entering (leaving) the graphene layer, $\lambda$ is the wavelength, and $\theta$ is the angle that incident light makes with z-axis (normal to the structure). Since our focus is to optimize structures for normal incidence of light, $\theta$ is set to be zero in optimizations.

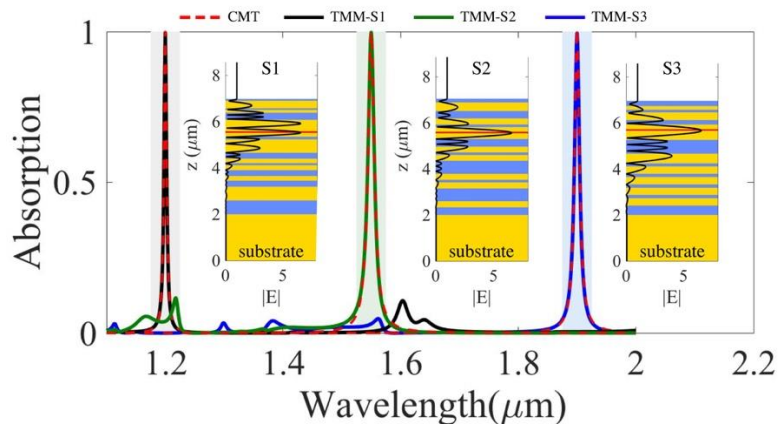

Fig. 2. Absorption spectra of three genetic-algorithm-optimized $M_6N_{12}$ structures (S1, S2, S3) for single-wavelength absorptions at 1.2 μm (black line), 1.55 μm (green line), and 1.9 μm (blue line). The results are compared with coupled mode

theory (CMT; dashed red line) demonstrating excellent fit near resonances. Note that CMT curves are disconnected and have been calculated individually for each peak. Corresponding E-field profiles of absorptions, shown as insets, illustrate strong enhancement of *E*-fields near graphene monolayer. Normalized *E*-field amplitudes at the graphene surface, for all cases, is approximately 6.5. Here, the silicon, silica, and graphene monolayer are shown by the blue, yellow, and red layers, respectively.

Using transfer matrix method and genetic algorithm as outlined, we first investigate the tunability properties of 1D aperiodic multilayers for single-wavelength light abortion. As shown in Fig. 2, when each structure (S1, S2, S3) is optimized for preselected absorption wavelengths, near total resonant light absorption happens by the graphene monolayer. Here, absorption wavelengths are set to be 1.2 $\mu$m (black line), 1.55 $\mu$m (green line), 1.9 $\mu$m (blue line) and $M_6N_{12}$ structure is used. As seen, desired wavelengths can be selected anywhere in the spectrum window and genetic optimization gives us proper aperiodic multilayers to ensure 100% light absorption. It should be mentioned that various aperiodic structures can be found for a specific wavelength selection; here, we have presented only one of them for each case.

Normalized E-field amplitudes profiles of absorption wavelengths, shown as insets of Fig. 2, indicate resonant enhancement of the E-fields at graphene surface for all three cases, which leads to near 100% light absorption in graphene monolayer. The E-fields' values at the graphene surface ($E_g$) are approximately the same: 6.57(S1), 6.58(S2), 6.56(S3). Since graphene is an ultra-thin material, it acts as a small perturbation to the system and does not disturb the E-field profiles of the resonance modes of the original structures (without graphene). However, it drops enhancement factors inside the structure and rescales the fields profile. For example, maximum values of E-fields ($E_{max}$) inside {S1, S2, S3} are {12.47, 13.15, 12.92} (without graphene) and {6.57, 6.6, 6.56} (with graphene), which shows that enhancement factors inside the structures has dropped by factor of two. In general, approximately equal amplitudes of E-fields at the graphene position for different wavelengths and the rescaled profiles of absorption modes over the wide range of wavelengths stem from a relatively small change of graphene loss; this indicates critical coupling of incoming waves to graphene monolayer.

In fact, an aperiodic multilayer with a graphene monolayer acts as a lossy resonating system with three possible states of under coupling, critical coupling, and over coupling. When the external leakage rate of resonance mode and intrinsic loss rates are equal, perfect absorption in graphene happens, and the system experiences critical coupling. This can be explained by CMT, which is based on input-output properties of a resonating system and coherent interference of direct and indirect pathways [14, 33]. Since the designed structures have ~100% light reflection at selected wavelengths, in the absence of graphene, they can be modeled as a one port resonating system. Therefore, the absorption spectrum of the graphene monolayer near absorption wavelengths for a multi-resonance system can be described as:

$$A(\omega) = \sum_{i=1}^{n} 4\delta_i \gamma_i / ((\omega - \omega_{0i})^2 + (\delta_i + \gamma_{ei})^2) \qquad (2)$$

where $\omega_{0i}$ is absorption resonance frequency, and $\gamma_{ei}$ and $\delta_i$ are external leakage rate and intrinsic loss rate at each resonance frequency, respectively. As Eq. 2 suggests, for a resonating structure with equal intrinsic loss and external leakage rates (critical coupling), the absorption will be one (or 100%). Non-equal rates cause partial absorption (over coupling/under coupling states). To validate this analysis, TMM simulations are compared with CMT results and demonstrated in Fig. 2 showing excellent agreement.

According to the above discussion, a small variation of graphene loss and a stable critical coupling to graphene in wide range of frequencies, will result in simultaneous tuning of multi-wavelength absorptions of the graphene monolayer and, also, similar $E$-field amplitudes at the graphene surface. To show this, we first optimize the $M_6N_{16}$ structures to have perfect double-wavelength light absorptions. Figs. 3(a) shows that when each structure is optimized for two preselected wavelengths of $\{\lambda_1 (\mu m), \lambda_2 (\mu m)\}$, near total resonant light absorption happens in the graphene monolayer at these wavelengths. Here, absorption wavelengths are set to be {1.5, 1.6} (blue line), {1.4, 1.7} (red line), {1.3, 1.8} (black line), and {1.2, 1.9} (green line). These preselected wavelengths are simultaneously tuned in a broad wavelength range extending from ~1.2 μm to ~1.9 μm. Corresponding $E_g$ of these wavelengths (from shortest to the longest) are 6.56, 6.46, 6.51, 6.54, 6.56, 6.58, 6.56, 6.55, respectively.

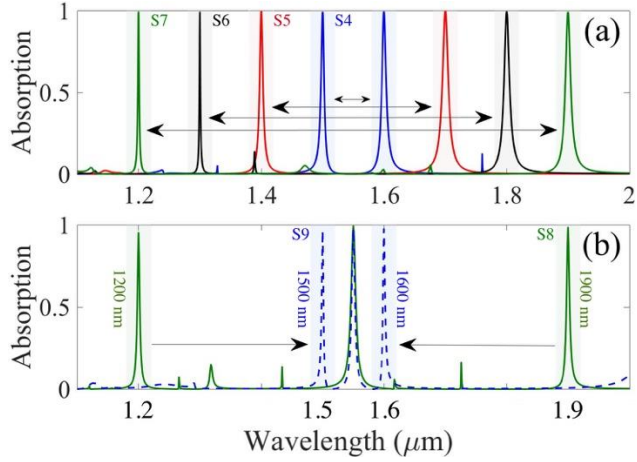

Fig. 3. Absorption spectra of genetic-algorithm-optimized structures as function of wavelength at normal incidence for preselected wavelengths (μm) of (a) **S4**: {1.5, 1.6}, **S5**: {1.4, 1.7}, **S6**: {1.3, 1.8}, **S7**: {1.2, 1.9} and (b) **S8**: {1.2, 1.55, 1.9}, **S9**: {1.5, 1.55, 1.6}. As seen, preselected wavelengths are tuned both *simultaneously* and *independently*. $M_6N_{16}$ ($M_6N_{18}$) arrangement is used for double- wavelength (triple-wavelength) absorptions.

For further exploration, we optimize $M_6N_{18}$ multilayers to absorb optical waves at three preselected wavelengths of {$\lambda_1$ (μm), $\lambda_2$(μm), $\lambda_3$(μm)}. In this case, one of the preselected absorption wavelengths ($\lambda_2$) is set to be 1.55 μm and other two wavelengths ($\lambda_1$, $\lambda_3$) independent from $\lambda_1$ are simultaneously tuned from {1.2, 1.9} to {1.5, 1.6}. As seen in Figs. 3(b), near perfect absorption of incident light is achieved at preselected wavelengths of {1.2,1.55,1.9} and {1.5,1.55,1.6}.

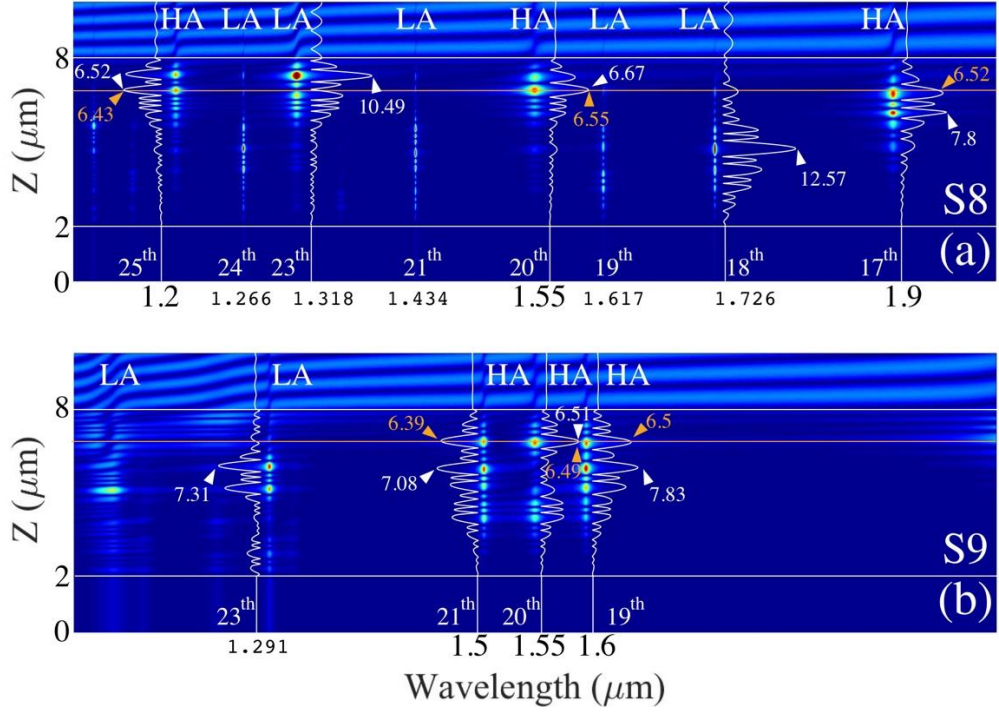

Fig. 4. Normalized *E*-field amplitudes λ-z map of **S8** and **S9** structures for absorption wavelength sets of {1.5, 1.55, 1.6} and {1.2, 1.55, 1.9}. As seen low-absorption resonance modes (LA-modes) are also excited in addition of high-absorption resonance modes (HA-modes). Also, orange arrows show $E_g$ of HA-modes. Modes' number are determined based on number of E-fields antinodes inside of structures.

In Fig. 4, we illustrate the λ-z map of normalized *E*-fields amplitude for the genetic-algorithm-optimized structures, S8 and S9, described in Fig. 3. Position of graphene monolayer and structure boundaries are shown by the orange and white lines. $E_{max}$ and $E_g$ are shown with white and orange arrows. Moreover, a 1D plot of *E*-fields of some of resonance modes of S8 and S9 statures are shown as insets (white). It is seen that two types of high-absorption modes (HA-modes: 17th, 20th, 25th for S8 and 19th, 20th, 21th for S9 and) and low-absorption modes (LA-modes: 17th, 22th for S8 and 23th for S9) coexist in the structures. These modes have different $E_{max}$ (white arrows), penetration depths, and frequency spreading. In S8, $E_{max}$ of the 18th is considerably larger than other modes which can be attributed to its relatively symmetric distribution of *E*-fields in the center of the structure and localization of *E*-fields far away from graphene monolayer. Critical coupling of 23th mode is prohibited and no absorption is observed which is related to the position of graphene monolayer on the node of *E*-field profile. Similar behavior is also seen for the S9 structure. Moreover, it is seen that, similar to previous cases, $E_g$ of HA-modes are similar and approximately equal to ~6.5 which, in general, works for any other optimized structures with perfect (near perfect) absorptions.

Table 1. Angular FWHM $\delta\theta_{TE/TM}$ (degree) and Spectral FWHM $\delta\lambda$ (nm) of the Structures described in Fig. 3

| Double-Wavelength Absorptions ($\mu$m) (Fig. 3(a)) | | | | |
|---|---|---|---|---|
| $\{\lambda_1, \lambda_2\}$ | {1.2, 1.9} | {1.3, 1.8} | {1.4, 1.7} | {1.5, 1.6} |
| $\{\delta\lambda_1, \delta\lambda_2\}$ | {2.6, 9.9} | {2.1, 10.3} | {6.2, 11.2} | {6.0, 8.1} |
| $\{\delta\theta_{TM1}, \delta\theta_{TM2}\}$ | {10.9, 15.1} | {10.2, 16.8} | {16.9, 19.3} | {12.2, 14.7} |
| $\{\delta\theta_{TE1}, \delta\theta_{TE2}\}$ | {10.3, 16} | {11.2, 18.0} | {16.9, 19.2} | {12.6, 14.6} |
| Triple-Wavelength Absorptions ($\mu$m) (Fig. 3(b)) | | | | |
| $\{\lambda_1, \lambda_2, \lambda_3\}$ | {1.2, 1.55, 1.9} | | {1.5, 1.55, 1.6} | |
| $\{\delta\lambda_1, \delta\lambda_2, \delta\lambda_3\}$ | {4.5, 8.9, 5.6} | | {3.4, 6.2, 4.1} | |
| $\{\delta\theta_{TM1}, \delta\theta_{TM2}, \delta\theta_{TM3}\}$ | {14.8, 17.2, 13.1} | | {10.8, 15.6, 13.4} | |
| $\{\delta\theta_{TE1}, \delta\theta_{TE2}, \delta\theta_{TE3}\}$ | {15.5, 17, 14.8} | | {11.2, 15.7, 12.1} | |

Angular aperture of incident light contributes to lowering of the maximal absorption since it means a deviation from the ideal 1D propagation problem. For comparison of the structures, we define the angular width at half maximum (FWHM), $\delta\theta$, calculating the width about $\theta = 0$ for which the absorbance is larger than half of the maximum achieved by genetic optimization for normal incident. We also define the spectral FWHM $\delta\lambda$ for the aperiodic structures, calculating the width about the absorption wavelengths at which we optimized the structures, for which the absorption is larger than half of the absorption values of the preselected wavelengths. Results are summarized in Table 1; as seen, both TE and TM modes have similar $\delta\theta$ and absorption process can be considered polarization independent around the normal incident. Moreover, absorptions with lower $\delta\lambda$ (higher Q-factors; Q= $\lambda/\delta\lambda$) have smaller $\delta\theta$ and show more angular sensitivity.

Finally, we investigate selectable multi-wavelength perfect absorptions by aperiodic structures in visible spectrum. Considering scalability of photonic structures, similar behavior applies to the visible wavelengths, and the difference is using proper lossless optical materials. In visible regime available range of index contrast is less than near-infrared and losses in silicon become considerable in visible range. As a result, we optimize aperiodic multilayers composed of alternating layers of tantalum pentoxide ($Ta_2O_5$) and silica deposited on a silica substrate for the wavelength range of 500 nm – 650 nm. $Ta_2O_5$, is lossless in this regime and has a refractive index of ~2.12.

As shown in Fig. 5, near perfect absorptions have been achieved for preselected wavelengths ($\mu$m) of {0.575}, {0.525, 0.625}, {0.525, 0.575, 0.625}. These wavelengths can be tuned both simultaneously and independently as discussed for the previous case. Genetic-algorithm-optimized structures for single-, double-, and triple-wavelengths absorptions are $M_{14}N_{26}$, $M_{18}N_{34}$, and $M_{18}N_{34}$, respectively. In this case, aperiodic structures have larger number of layers compared to the near-infrared structures which is due to the low index contrast of $Ta_2O_5$ and $SiO_2$ in comparison with Si and $SiO_2$. In fact, the ratio of $n_H/n_L$ for the first one is ~1.46 and for the second one is ~2.4.

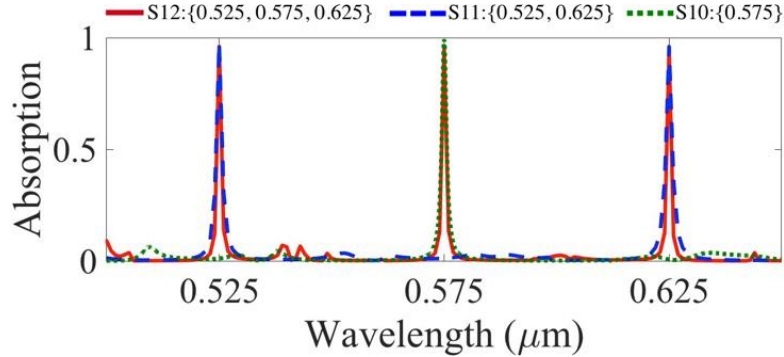

Fig. 5. Absorption spectra of genetic-algorithm-optimized structures as function of wavelength at normal incidence for preselected wavelengths ($\mu$m) of **S10**: {0.575}, **S11**: {0.525,0,625}, **S12**: {0.525, 0.575, 0.625}.

In summary, we have demonstrated that 1D aperiodic multilayer structure can be successfully optimized to achieve near total absorption in a graphene monolayer at preselected wavelengths. These absorption peaks are highly tunable and can be controlled either simultaneously or independently. Simulation results reveals that normalized *E*-fields value at graphene surface is always ~6.5 when perfect (near perfect) absorption happens. Proposed platform is a promising candidate to design multispectral light detection devices and filters; it is metal free and do not require surface texturing or patterning. Presented concept is applicable for other 2D materials.

**ACKNOWLEDGEMENTS**

Authors would like to acknowledge support from the Army Research Office and the Defense Advanced Research Projects Agency. We are also grateful for the support of Air Force Office of Scientific Research (AFOSR) Small Business Innovation Research (SBIR) program under award number FA9550-17-P-0014.